\documentclass[12pt,a4paper]{article}

\usepackage{amsmath,amssymb, amsfonts, amsthm}

\title{Classical random field viewpoint to Gorini-Kossakowski-Sudarshan-Lindblad equation and its linear and nonlinear 
generalizations}

\author{International Center for Mathematical Modeling\\
 in Physics and Cognitive Sciences \\
 Linnaeus University, S-35195, V\"axj\"o, Sweden}

\begin{document}

\maketitle

\begin{abstract}    
We show that the basic equation of theory of open systems, Gorini-Kossakowski-Sudarshan-Lindblad equation, as well
as its linear and nonlinear generalizations
have a natural classical probabilistic interpretation -- in the framework of prequantum classical statistical field 
theory. The latter gives an example of a classical probabilistic model (with random fields as subquantum variables)
reproducing the basic probabilistic predictions of quantum mechanics.   
\end{abstract}

\section{Introduction}

Recently interest to probing the limits  of quantum mechanics essentially increased -- both 
quantum contra classical and quantum contra ``superquantum'' models (i.e., models deviating from classical physics even more than it does ``conventional quantum mechanics''), see, e.g., \cite{Z}--\cite{K3}. On one hand, this situation is a consequence 
of the internal development of physical science. On the other hand, it became clear that the projects of 
quantum computing and quantum cryptography have some difficulties in realization and new deep foundational analysis 
is needed. In particular, the role of environment cannot be ignored and, instead of unitary dynamics, open systems
dynamics (including non-Markovian) may play an important role in quantum computing and other applications. In this 
paper  we present the quantum contra classical probing of open system dynamics; in particular,
Gorini-Kossakowski-Sudarshan-Lindblad (GKSL)equation. The result of our consideration 
does not match with the rather common expectation. We show that, for any semigroup of positive (in particular, completely positive) maps, the density operator dynamics can be interpreted in the purely classical way. We derive the corresponding differential equation 
for the classical probabilistic dynamics. It is very complicated comparing with the original linear operator dynamics: in general
nonlinear and integro-differential. Therefore operationally it is natural to work the standard quantum dynamical equation.
However, the possibility to interpret this equation in the classical probabilistic terms has nontrivial foundational consequences.

The study in this paper has the natural coupling to {\it prequantum classical statistical field 
theory} (PCSFT), see, e.g., \cite{K0}, \cite{K1}, \cite{K2}, \cite{K3}, \cite{K1a}. The latter gives an example of a classical probabilistic model, with random fields as subquantum variables,
reproducing the basic probabilistic predictions of quantum mechanics. The basic idea beyond PCSFT is very simple: to couple 
quantum states, density operators, with the covariance operators of classical (``prequantum'') random fields in complex Hilbert spaces. Of course,
the covariance operator does not determine a random field uniquely. However, by restricting prequantum fields to Gaussian 
fields we determine the prequantum field by its covariance operator. (We consider 
fields with zero average.) Of course, in general the covariance operator of a random field has non-unit trace. Therefore
on the subquantum level there is no reason to consider the trace-preserving dynamics. And we proceed in this paper with 
semigroups of positive maps which in general  are not trace preserving.
In section \ref{ZZTT} we derive the dynamical equation for density operators corresponding to such dynamics of covariance operators.
This equation is nonlinear and it has quadratic nonlinearity.

\section{ Prequantum Classical Statistical Field
Theory} \label{PCSFT}

 Here we briefly present essentials of PCSFT, see \cite{K0}, \cite{K1}, \cite{K2}, \cite{K3}, \cite{K1a} for detailed presentation. 
To simplify considerations, we will study quantum systems with
finite dimensional Hilbert spaces. 

Take complex Hilbert space $H$ as space of classical ``prequantum'' states.  Consider a probability
distribution $\mu$ on  $H$  having zero average (it means
that $\int_H \langle \phi, y \rangle  d\mu(\phi)=0$ for any $y \in H$) and the
covariance operator $B$ which is  defined by the Hermitian positively defined bilinear form:
\begin{equation}
\label{CO} \langle  B y_1, y_2 \rangle =
\int_H \langle y_1, \phi\rangle \langle \phi,  y_2\rangle d\mu(\phi),
y_1, y_2 \in H,
\end{equation}
By scaling we obtain the operator
\begin{equation}
\label{COPI}
 \rho\equiv \rho_B = B/\rm{Tr} B,
\end{equation}
 with $\rm{Tr} \rho=1.$
Mathematically it has all properties of the density operator.

By PCSFT a quantum state, a density operator, is simply the
symbolic representation  of the covariance operator  of the
corresponding prequantum (classical) probability distribution.

In general a probability distribution  is not
determined by its covariance operator. Thus the correspondence
PCSFT $\to$ QM is not one-to-one. However, if the class of
prequantum probability distributions  is restricted to Gaussian,
then this correspondence becomes one-to-one. 

In PCSFT classical
variables are defined as functions from (Hilbert) state space $H$
to real numbers, $f=f(\phi).$ By PCSFT a quantum observable, a
Hermitian  operator, is simply a symbolic representation of $f$
by means of its second derivative (Hessian), $f \to
\widehat{A}=\frac{1}{2} f^{\prime \prime}(0).$ This correspondence
is neither one-to-one. However, by restricting the class of
classical variables to quadratic forms on Hilbert state space $H,$
$f_A(\phi) =\langle \widehat{A} \phi, \phi \rangle,$    we make  correspondence
PCSFT $\to$ QM one-to-one. And finally, we present the basic
equality coupling prequantum (measure-theoretic)  and quantum (operator) averages: 
\begin{equation}
\label{CO1} \langle f_A\rangle_\mu\equiv E f_A  =\int_H
f_A(\phi)  d\mu(\phi)= \rm{Tr} B \widehat{A}=
 (\rm{Tr} B) \;  \langle \widehat{A} \rangle_\rho,
\end{equation}
where 
\begin{equation}
\label{CO7}
\langle \widehat{A} \rangle_\rho = \rm{Tr} \rho \widehat{A}.  
\end{equation}
Thus the quantum average $\langle \widehat{A} \rangle_\rho$ can be
obtained as just scaling of the classical average. We remark that
scaling parameter $\rm{Tr} D$ is, in fact, the dispersion of the
probability distribution $\mu:$
\begin{equation}
\label{TR} \sigma^2\equiv E||\phi||^2 =
 \int_H ||\phi||^2 d\mu(\phi)={\rm Tr}\; B.
\end{equation}
It describes the strength of deviations of a random vector $\phi$
from its average. The latter is zero in our model. If $H=L_2({\bf R}^m),$ the space of square integrable 
functions, $\phi : {\bf R}^m \to {\bf C},$ then
we can consider $\phi$ as a random field: $\phi=\phi(x, \omega),$ where $x \in {\bf R}^m$ and 
$\omega$ is the chance parameter.
The average is the field which is equal to zero everywhere. Suppose that
$\sigma^2$ is very small, so deviations from zero  are small.
Since
\begin{equation}
\label{CO2}\langle \widehat{A} \rangle_\rho = \frac{1}{\sigma^2}
\langle f_A\rangle_\mu,
\end{equation}
the quantum average can be considered as the average of amplified
``prequantum signal''.

{\bf Remark.} The basic relation of PCSFT coupling quantum states (density operators) with covariance operators 
of prequantum processes, see (\ref{COPI}), can be used the other way around. M. Ohya and N. Watanabe \cite{W} 
used it in classical signal theory to define information (in the case of Gaussian noisy channels with the infinite 
numeber of degrees of freedom) with the aid of von Neumann entropy.  

\section{Dynamics of the probability measure corresponding to semigroup of positive maps} 
\label{HUHU}

Consider a semigroup of positive maps $C_t= e^{t {\cal L}}$ which is induced by the generator  given by 
the superoperator ${\cal L}.$ Although the general form of the generator ${\cal L}$ is unknown, there are 
known numerous examples of such generators. For examples, the general form of the generator of the completely positive 
semigroup was described in the relation with the GKSL-equation. The latter equation gives the most 
important (for applications) example of the dynamics under consideration. 
  We set
$B_t= C_t B_0,$ where $B_0$ is a positive operator Hermitian operator. Then $B_t$ is a family of positive operators giving the solution of 
the Cauchy problem:
\begin{equation}
\label{GGHH}
\frac{d B_t}{dt} = {\cal L} B_t, B_{t=0} = B_0.
\end{equation}
Let us denote by $P_t$ the Gaussian measure with zero mean value and the covariance operator $B_t.$ 
Our aim is to derive the differential equation for the map $t\to P_t.$ To simplify the derivation,
we consider the case of finite dimensional Hilbert space. Here $P_t$ has the density 

\begin{equation}
\label{GGHH1}
 p_t(z, z^*) =\frac{1}{\sqrt{(2\pi)^n \rm{det} B_t}} e^{-  \langle B_t^{-1} z, z\rangle}.
\end{equation} 
By selecting some orthonormal basis $(e_j),$ i.e., $z=\sum_j z_j e_j, z_j= \langle z, e_j\rangle,$
we represent the density as
\begin{equation}
\label{GGHH2}
 p_t(z, z^*) =\frac{1}{\sqrt{(2\pi)^n \rm{det} B_t}} e^{-  \sum_{ij} \langle B_t^{-1} e_i, e_j\rangle z_i  z_j^*}.
\end{equation} 
This density can be represented by means of its Laplace-Fourier transform as 
\begin{equation}
\label{GGHH3}
 p_t(z, z^*) =\int   e^{- \langle B_t \xi, \xi \rangle + 
i \langle \xi, z\rangle+  i \langle z, \xi\rangle} d \xi d \xi^*
\end{equation}
or in the coordinate form:
\begin{equation}
\label{GGHH3A}
 p_t(z, z^*) =\int   e^{-  \sum_{ij}\langle B_t e_i, e_j\rangle \xi_i \xi_j^* + 
i \sum_j (\xi_i z_i^* + \xi_i^* z_i)} d\xi d \xi^* 
\end{equation}
We have 
$$
\frac{d p_t}{d t}(z,  z^*) = -   \sum_{ij}\langle \frac{dB_t}{dt} e_i, e_j\rangle 
\int   \xi_i \xi_j^* e^{- \sum_{ij}\langle B_t e_i, e_j\rangle \xi_i \xi_j^* + i 
\sum_j (\xi_i z_i^* + \xi_i^* z_i)} d\xi d \xi^*
$$ 
$$
= \sum_{ij}\langle {\cal L} B_t e_i, e_j\rangle \frac{\partial^2}{\partial z_j \partial z_i^*} 
\int   e^{-  \sum_{ij}\langle B_t e_i, e_j\rangle \xi_i \xi_j^* + i
\sum_j (\xi_i z_i^* + \xi_i^* z_i)} d\xi d \xi^*
$$
$$
= \sum_{ij}\langle   B_t,  {\cal L}^* (\vert e_i\rangle \langle e_j\vert )\rangle  \frac{\partial^2}{\partial z_j \partial z_i^*} p_t(z, z^*),
$$
where in the last expression we used the scalar product in the space of Hilbert-Schmidt operators.
We can expand $B_t$ with respect to the orthonormal basis $(\vert e_k\rangle \langle e_m\vert),$
$$
B_t = \sum_{km} \langle B_t e_k, e_m\rangle \vert e_k\rangle \langle e_m\vert.  
$$
Finally, we have
\begin{equation}
\label{GGHH3AB}
\frac{d p_t}{d t}(z, z^*) = \sum_{km; ij} {\cal L}_{km, ij} \int u_k u_m^* p_t(u, u^*) du d u^* 
\frac{\partial^2}{\partial z_j \partial z_i^*} p_t(z, z^*),
\end{equation}
where 
\begin{equation}
\label{GGHH3ABc}
{\cal L}_{km, ij}  = \langle {\cal L} \vert e_k\rangle \langle e_m\vert, \vert e_i \rangle \langle e_j\vert\rangle 
\end{equation}
By using notation from statistical mechanics we write our equation as 
\begin{equation}
\label{GGHH3ABd}
\frac{d p_t}{d t}(z, z^*) = \sum_{km; ij} {\cal L}_{km, ij}  \overline{z_k z_m^*}  \frac{\partial^2}{\partial z_j \partial z_i^*} p_t(z, z^*).
\end{equation}
We now show that, for any Gaussian solution of the latter equation, its covariance operator satisfies 
the equation (\ref{GGHH}).   We have 
$$
\frac{d }{dt} \langle B_t e_i, e_j\rangle= \int z_i z_j^* \frac{dp_t}{dt} (z, z^*) d z d z^* =
$$
$$  
 - \int z_i z_j^* \Big( \int \langle {\cal L} B_t \xi, \xi\rangle e^{- \langle B_t \xi, \xi \rangle + 
i \langle \xi, z\rangle+  i \langle z, \xi\rangle} d \xi d \xi^* \Big)  d z d z^*
$$
$$
=  -\int \langle {\cal L} B_t \xi, \xi\rangle  \Big( \int z_i z_j^* e^{i \langle \xi, z\rangle+  i \langle z, \xi\rangle}  d z d z^* \Big) e^{- \langle B_t \xi, \xi \rangle} d \xi d \xi^*  
$$
$$
=  \int \langle {\cal L} B_t \xi, \xi\rangle  \frac{\partial^2}{\partial \xi_j \xi_i^*}\Big( \int  e^{i \langle \xi, z\rangle+  i \langle z, \xi\rangle}  d z d z^* \Big) e^{- \langle B_t \xi, \xi \rangle} d \xi d \xi^*
$$
$$
= \int \langle {\cal L} B_t \xi, \xi\rangle  \frac{\partial^2 \delta}{\partial \xi_j \xi_i^*} (\xi, \xi^*) e^{- \langle B_t \xi, \xi \rangle} d \xi d \xi^*
= \langle {\cal L} B_t e_i, e_j\rangle.
$$
Thus any dynamics of the form (\ref{GGHH}) in the space of positive operators  
can be imagined as corresponding to the dynamics (\ref{GGHH3ABd}) in the space 
of Gaussian measures. The latter is nonlinear and it is integro-differential. 
Of course, the former is essentially simpler than the latter. Therefore one can 
restrict considerations to the dynamics of operators and proceed in the 
phenomenological framework, i.e., without coupling these operators to Gaussian 
measures (``distributions of hidden parameters'').

{\bf Example 1.} Consider our equation for the probability density in the real case and for dimension one. It has 
the form:
\begin{equation}
\label{LULU}
\frac{d p_t}{dt}(x)= \frac{A}{2} \bar{x^2} p_t(x).
\end{equation}
(the factor 1/2 is related to consideration of the real case).
Consider the solution of the form
$$p_t(x) = \frac{e^{-\frac{x^2}{2B_t}}}{\sqrt{2 \pi B_t}}.$$ Then
$$\frac{d p_t}{dt}(x)=\frac{\frac{d B_t}{dt}}{2} \Big( \frac{x^2}{B_t^2} - \frac{1}{B_t}\Big) p_t(x)$$
and $$p_t^{\prime\prime}(x)=\Big( \frac{x^2}{B_t^2} - \frac{1}{B_t} \Big) p_t(x).$$ The equation (\ref{LULU}) implies
that      $\frac{d B_t}{dt}= A B_t,$ the covariance (in this case simply dispersion) evolves linearly.

{\bf Example 2.} Consider the dynamics:
\begin{equation}
\label{LULU1}
\frac{dB_t}{dt} = A B_t + B_t A^*, B_{t=0}= B_0, 
\end{equation}
where $A$ is some operator (in general non-Hermitian). Then 
\begin{equation}
\label{LULU7}
B_t= e^{At} B_0 e^{A^*t}.
\end{equation} 
Consider now the dynamics of the ``state vector'' with a random initial condition 
\begin{equation}
\label{LULU8}
\frac{d\phi_t}{dt} (\omega) = A \phi_t(\omega), \phi_{t=0}(\omega) = \phi_{0}(\omega), 
\end{equation}
where $\phi_0$ has the Gaussian distribution with zero mean value and the covariance operator $B_0.$ 
Its solution $\phi_t(\omega) = e^{At}\phi_{0}(\omega)$
has the covariance operator given by (\ref{LULU7}). 
At the same time we know from theory of classical stochastic processes that the probability distribution
of $\phi_t$ satisfies the trivial forward Kolmogorov equation with zero diffusion. This equation
is linear. Thus in some special cases nonlinearity and integro-differential structure of the 
equation for the Gaussian probability distribution corresponding to the linear dynamics of the covariance
operator are redundant. In particular, set  $A= - i H,$ where $H$ is the hermitian operator. Then the equation
(\ref{LULU1}) is nothing else than the von Neumann equation, see also \cite{K1}:
\begin{equation}
\label{LULU77}
\frac{dB_t}{dt} =-i [H,B_t], B_{t=0}= B_0, 
\end{equation}
  
\section{Nonlinear dynamics of the density operator}
\label{ZZTT}

Our aim was representation of quantum mechanics with the aid of classical 
Gaussian distributions. In general the trace of a covariance operator is not equal 
to one. Moreover, the dynamics considered in the previous section need not preserve the trace,
i.e., even by starting with a density operator $\rho_0\equiv B_0,$ we need not obtain a trajectory 
in the space of quantum states.

To obtain the density operator corresponding to the covariance operator of the prequantum random field, we normalize the covariance operator of a Gaussian measure by its trace:
\begin{equation}
\label{GGHH3Aq}
B_t \to \rho_t=\frac{B_t}{\rm{Tr} B_t}.
\end{equation}
Then the differential equation (\ref{GGHH}) is transformed to a differential equation for $\rho.$ 
$$
\frac{d \rho_t}{dt}= \frac{\frac{d B_t}{dt}}{\rm{Tr} B_t}- \frac{B_t \rm{Tr} \frac{ d B_t}{dt}}{\rm{Tr}^2 B_t}
$$
$$
=\frac{{\cal L} B_t}{\rm{Tr} B_t} -       \frac{B_t \rm{Tr} {\cal L} B_t }{\rm{Tr}^2 B_t}= 
{\cal L} \rho_t - \rho_t \rm{Tr} {\cal L} \rho_t.
$$ 
Thus the  evolution equation for the density operator corresponding to the covariance operator is nonlinear 
with quadratic nonlinearity:
\begin{equation}
\label{GGHH3Ar}
\frac{d \rho_t}{dt} = {\cal L} \rho_t - \rho_t \rm{Tr} {\cal L} \rho_t.
\end{equation}
Suppose that the original covariance-dynamics is trace preserving. Than we have:
 $$
0= \frac{d \rm{Tr} \rho_t}{dt}= \rm{Tr} \frac{d \rho_t}{dt} = \rm{Tr} {\cal L} \rho_t.$$
Thus the last term in (\ref{GGHH3Ar}) disappears and dynamics for the covariance and density 
operators coincide and they are both linear:
\begin{equation}
\label{GGHHppp}
\frac{d \rho_t}{dt} = {\cal L} \rho_t, \rho_{t=0} = \rho_0.
\end{equation}
However, from the prequantum viewpoint there is no reason to consider only trace preserving dynamics of 
covariance operators. Hence, the basis dynamical equation is   the nonlinear equation (\ref{GGHH3Ar}). 

\section{Brownian motion and Ornstein-Uhlenbeck processes}

Consider the Brownian motion in $H.$ It is determined by its covariance operator $\Sigma$ (Hermitian and positively defined).
The process $w_\Sigma(t, \omega)= \xi_0(\omega) + \sqrt{\Sigma} w(t,\omega),$ where $w(t)$ is the standard Brownian motion, i.e.,   
$w(t) \sim {\cal N} (0, t I),$ where $I$ is the unit operator and $\xi_0$ is a random variable. Suppose that the latter is independent 
from $w(t);$ denote its covariance operator by $B_0.$ Then the covariance operator $B_t$ of $w_\Sigma(t)$ satisfies to the equation:
\begin{equation}
\label{GGHHppp_77}
\frac{d B_t}{dt} = \Sigma, \; \;  B_{t=0} = B_0.
\end{equation}
The corresponding ``density operator'' has the form $$
\rho_t= \frac{B_0 + \Sigma t}{\rm{Tr} (B_0 + \Sigma t)}.
$$ 
This function
satisfies to the ugly differential equation with time dependent coefficients. Thus the classical Wiener process 
is not among processes corresponding to semi-groups of positive linear maps.

Now consider the solution of stochastic differential equation
$$
d \phi_t(\omega) = A_t \phi_t(\omega) dt + d w_\Sigma(t,\omega),\;  \phi_0(\omega) = \xi_0(\omega),
$$  
where the initial condition is again independent from the $w_\Sigma(t).$ (In particular, this type of equations describes 
the Ornstein-Uhlenbeck process). 
Then it is easy to show (by using the Ito formula) that its covariance operator $B_t$ satisfies the equation:   
\begin{equation}
\label{GGHHppp_77}
\frac{d B_t}{dt} = A_t B_t + B_t A_t^* + \Sigma, \; \; B_{t=0} = B_0.
\end{equation}
Thus the classical Ornstein-Ulenbeck process is neither in the club of ``prequantum processes.'' 

By selecting the time independent coefficient $A_t= - i {\cal H},$ where ${\cal H}$ is a Hermitian operator, we obtain the 
dynamics: 
\begin{equation}
\label{GGHHppp_77}
\frac{d B_t}{dt} = -i [H,B_t] + \Sigma, \; \; B_{t=0} = B_0.
\end{equation}
This equation is just inhomogeneous von Neumann equation. (It might be that such equations were used in QM, but 
I do not know such applications.) 

This paper was prepared during the visit to the Center for Quantum Bio-Informatics (Tokyo University of Science), 
February-March 2013. The author would like to thank Massanori Ohya and Noboru Watanabe for fruitful discussions and 
hospitality.


\begin{thebibliography}{400}

\bibitem{Z} K. Zyczkowski, Quartic quantum theory: an extension of the standard quantum mechanics.
{\it J. Phys. A: Math. Theor.}  (2008) {\bf 41} 355302.

\bibitem{Sorkin}  Sinha, U., Couteau, C., Medendorp, Z.,
Sollner, I., Laflamme, R.,  Sorkin,  R., and  Weihs, G.:
Testing Born's rule in quantum mechanics with a triple slit
experiment. In:  Accardi, L., Adenier, G., Fuchs, C., Jaeger, G.,
Khrennikov, A., Larsson, J.-A., and Stenholm, S. (eds).
Foundations of Probability and Physics-5, , pp. 200-207.
American Institute of Physics, Ser. Conference Proceedings, vol.
1101. Melville, NY (2009).

 \bibitem{K1} A. Khrennikov,  Prequantum classical statistical field theory: Schr\"odinger dynamics of
entangled systems as a classical stochastic process, {\it  Found.
Phys.}   (2010) 1-13.

\bibitem{TH2} 't Hooft, G.: The free-will postulate in quantum
mechanics. {\it Herald of Russian Acad. Sc.} {\bf 81}  (2011)  907-911.

\bibitem{Groessing} G. Groessing,  
The vacuum fluctuation theorem: Exact Schr\"odinger equation via nonequilibrium thermodynamics.
{\it Phys. Lett.} A  {\bf 372}  (2008) 4556-4563.

\bibitem{Groessing1} G. Groessing, (ed.):    Emergent Quantum
Mechanics (Heinz von Foerster Congress). {\it J. Phys.: Conf. Ser.} (2012).

\bibitem{K2} A. Khrennikov, M. Ohya, N.  Watanabe,  Quantum probability from classical signal theory.
{\it International J. of Quantum Information} {\bf 9}  (2011) 281-292.

\bibitem{K3} A. Khrennikov, M. Ohya, N. Watanabe: Classical signal model from
quantum channels. {\it J. Russian Laser Research} {\bf 31}  (2010)
462-468.


\bibitem{K0} A.   Khrennikov, A pre-quantum classical statistical model
with infinite-dimensional phase space. {\it J. Phys. A: Math.
Gen.} {\bf 38}  (2005) 9051-9073.

 \bibitem{K1a} A. Khrennikov, Quantum mechanics as an approximation of
statistical mechanics for classical fields, {\it Rep. Math.
Phys.} {\bf 60}  (2007) 453-484.

\bibitem{W}  M.  Ohya, and N. Watanabe,
A new treatment of communication processes with Gaussian channels,
{\it Japan J. Industrial and Appl. Math.} {\bf 3} (1986) 197-206.

\end{thebibliography}
\end{document}